# Applications of the Differential Calculus in the Study of the Timed Automata: the Inertial Delay Buffer

Serban E. Vlad
the Computers Department of the Oradea City Hall,
p-ta Unirii, Nr. 1, 3700, Oradea, Romania
serbanvlad@excite.com, www.oradea.ro/serban/index.html

**Abstract** We write the relations that characterize the simpliest timed automaton, the inertial delay buffer, in two versions: the non-deterministic and the deterministic one, by making use of the derivatives of the $R \to \{0,1\}$ functions.

## 1. Introduction

The published literature in modeling the digital circuits from electrical engineering is rich and, facing it, the author proposes the next joke: *"Hey, friends, what about modeling the identity? the inertial delay buffer, the device that makes 0 be associated with 0 and 1 be associated with 1?"* We hope that our answer be not considered trivial, as well as it differs from certain answers that we have found. Moreover, because we use a certain language, we wish that our work shows a natural frame, that of the pseudoboolean differential calculus, in the analysis of the digital circuits.

The paper is organized as follows. First we define the signals and their derivatives. Then an informal definition of the inertial delay buffer is given, identical with the usual one, followed by a formal definition, the main result. Eventually, a comparison is made with literature.

## 2. Preliminaries

**2.1 Definition** $B = \{0,1\}$ is endowed with the order $0 \leq 1$, the discrete topology and the usual laws: the *complement* ' $\overline{\phantom{x}}$ ', the *product* ' $\cdot$ ', the *reunion* ' $\cup$ ', the *modulo 2 sum* ' $\oplus$ ', etc.

**2.2 Definition** The order and the laws from $B$ induce an order and laws on the set of the $R \to B$ functions. We keep the same names and notations.

**2.3 Definition** The next numbers are defined for the function $x : R \to B$ and $A \subset R$:

$$\bigcup_{\xi \in A} x(\xi) = \begin{cases} 1, \exists \xi \in A, x(\xi) = 1 \\ 0, else \end{cases}$$

$$\bigcap_{\xi \in A} x(\xi) = \begin{cases} 0, \exists \xi \in A, x(\xi) = 0 \\ 1, else \end{cases}$$

If $A$ is an interval of the form $(t-d,t), [t-d,t)$ where $d > 0$ is a parameter and $t$ runs in $R$, then the previous relations associate to $x$ respectively $R \to B$ functions.

**2.4 Definition** The non-negative strictly increasing unbounded sequence $0 = t_0 < t_1 < t_2 < ...$ is called *timed sequence*.

**2.5 Definition** We call *signal* or *realizable function* a function $x : R \to B$ with the properties:
   a) $\forall t < 0, x(t) = 0$

b) the timed sequence $(t_k)$ exists so that $\forall k \in N, x_{|[t_k, t_{k+1})}$, the restriction of $x$ at the interval $[t_k, t_{k+1})$, is constant.

The set of the signals is noted with $S$.

**2.6 Remark** Several timed sequences $(t_k)$ are associated with a signal $x \in S$ so that 2.5 b) be true. As a special case, if $x$ is the null function, any timed sequence makes 2.5 b) be true.

**2.7 Remark** It is easily shown that if $(t_k)$ is a timed sequence, then for any $t' < t$, the set $\{k \mid t' < t_k < t\}$ is finite (possibly empty). This property is related with what is called in the literature 'non-zeno' signals. Our definition 2.5 makes any signal be non-zeno.

**2.8 Definition** The *left limit* $x(t-0)$ and the *left derivative* $Dx(t)$ of the arbitrary function $x : R \to B$ *in the point* $t \in R$ are the binary numbers defined like this:
$$\exists t' < t, x_{|(t',t)} = x(t-0) \text{ (the constant function equal with } x(t-0)\text{)}$$
$$Dx(t) = x(t-0) \oplus x(t) = \overline{x(t-0)} \cdot x(t) \cup x(t-0) \cdot \overline{x(t)}$$
The numbers $\overline{x(t-0)} \cdot x(t)$, $x(t-0) \cdot \overline{x(t)}$ are called *left semi-derivatives in* $t$. When $t$ runs in $R$, the previous numbers associate with $x$ respectively $R \to B$ functions, having the same names and notations: the *left limit function* of $x$, $x(t-0)$ etc.

**2.9 Remark** An arbitrary function $x : R \to B$ may have or may have not left limit in some $t$. The signals $x \in S$ have left limit in any $t$ and consequently left derivative in any $t$, but generally the functions $x(t-0), Dx(t)$ are not signals.

**2.10 Example** The signal
$$x(t) = \begin{cases} 1, \text{ if } t \in [0,1) \vee [2,3) \\ 0, \text{ else} \end{cases}$$
has the timed sequence $(t_k) = N$ so that 2.5 b) is true. Moreover
$$x(t-0) = \begin{cases} 1, \text{ if } t \in (0,1] \vee (2,3] \\ 0, \text{ else} \end{cases}$$
$$Dx(t) = x(t-0) \oplus x(t) = \begin{cases} 1, \text{ if } t \in \{0,1,2,3\} \\ 0, \text{ else} \end{cases}$$

By comparing $x(t)$ with $Dx(t)$, we see how the derivative shows the moment when the function switches (from 0 to 1 or from 1 to 0). In other words, the support of the derivative coincides with the set of the (left) discontinuity points.

**2.11 Remark** The way that we have defined the signals, as right continuous functions having left discontinuities, makes us use left derivatives (the right derivatives being null) and refer to non-anticipative systems. The dual situation, of the $R \to B$ functions that are constant on intervals of the form $(t_{k+1}, t_k]$, is the one of the anticipative systems.

**2.12 Remark** We relate the present formalism with the usual computer science terminology like this: for $x \in S$ and $t \in R$, $x(t-0)$ is the 'old value' and $x(t)$ is the 'new value' of $x$ in $t$.

### 3. The Inertial Delay Buffer, Informal Aspects

**3.1 The informal definition of the non-deterministic inertial delay buffer** (NIDB). NIDB is a name given to two distinct things: an electrical device and a timed automaton, its mathematical model. In this paper we shall refer only to the latter, with the graphical symbol in the next figure

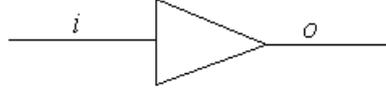

where $i, o$ are signals called *input* and *output*.

We define informally the automaton by its behavior and, in order to make a choice, we suppose that at the time moment $t_1 \geq 0$ the output is null and the input switches from 0 to 1. The real constants $0 < d_{r,\min} \leq d_{r,\max}$ are given so that the behavior of $o$ is indicated in the next table

| The switch of $o$ from 0 to 1 $o(t_1) = 0, i(t_1 - 0) = 0, i(\xi) = 1, \xi \in [t_1, t)$ | |
|---|---|
| $t \in [t_1, t_1 + d_{r,\min})$ | $o(t) = 0$ necessarily |
| $t \in [t_1 + d_{r,\min}, t_1 + d_{r,\max})$ | $o(t) = 0$ and $o(t) = 1$ are both possible |
| $t = t_1 + d_{r,\max}$ | if $o(t_1 + d_{r,\max} - 0) = 0$ then $o(t_1 + d_{r,\max}) = 1$ necessarily |
| $t > t_1 + d_{r,\max}$ | $o(t) = 1$ necessarily |

thus the next implication is true

$$\overline{o(t_1)} \cdot \overline{i(t_1 - 0)} \cdot \bigcap_{\xi \in [t_1, t_1 + d_{r,\max})} i(\xi) \leq \bigcup_{t \in [t_1 + d_{r,\min}, t_1 + d_{r,\max}]} \overline{o(t-0)} \cdot o(t)$$

The sense of $d_{r,\min}, d_{r,\max}$ is that of '*delay*' and '*raise*', from 0 to 1, of $o$.

There obviously exists a dual table of the previous one, corresponding to the '*fall*' of $o$ from 1 to 0 and characterized by the real constants $0 < d_{f,\min} \leq d_{f,\max}$. There also exists the possibility that $i$ changes its value before succeeding to produce a switch of the output; in intervals where $i(t) = o(t)$, the automaton remains indefinitely long (the property of *stability*).

Another possibility of describing the behavior of NIDB is given in the next figure.

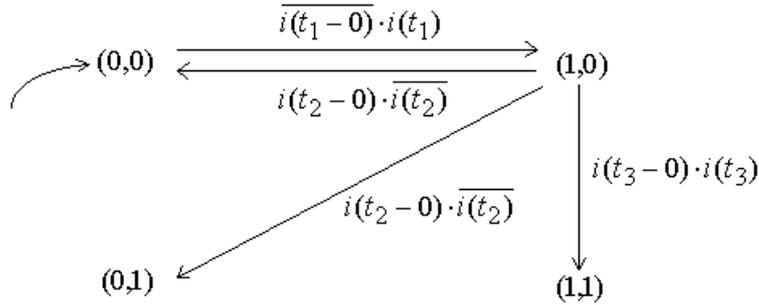

The couples $(i(t), o(t))$ are called the *states* of the automaton and they will be understood as points from $\boldsymbol{B}^2$ (not as functions $\boldsymbol{R} \to \boldsymbol{B}^2$).

The left arrow represents the initialization of the automaton. The fact that the initial state is $(0,0)$ is simply a choice that does not affect the generality.

$t_1 < t_2, t_3$ represent three values of the temporal variable $t$. We ask that in the intervals $[t_1, t_2), [t_1, t_3)$ the signals $i, o$ are constant.

The arrows represent the transitions of the automaton from one state in another state, at one of the time moments $t_1, t_2, t_3$ and the conditions $\overline{i(t_1 - 0)} \cdot i(t_1), i(t_2 - 0) \cdot \overline{i(t_2)}$, $i(t_3 - 0) \cdot i(t_3)$ are the causes that determine these transitions: the raise of the input from 0 to 1, the fall of the input from 1 to 0, respectively no change in the input value that remains 1.

We have not drawn in the figure the dual situations.

The behavior of the automaton is the following.

We suppose that at the time instant $t_1 \geq 0$ we have $o(t_1) = 0$ and $\overline{i(t_1 - 0)} \cdot i(t_1) = 1$ and this can happen for example if we take $t_1 = 0$ and the automaton is in the initial state. The transition $(0,0) \to (1,0)$ is thus *necessary* at $t_1$ and we have the next possibilities:

a) $i(t_2 - 0) \cdot \overline{i(t_2)} = 1$

- if $t_2 \in (t_1, t_1 + d_{r,\min})$ the transition $(1,0) \to (0,0)$ is *necessary* at $t_2$
- if $t_2 \in [t_1 + d_{r,\min}, t_1 + d_{r,\max})$ the transitions $(1,0) \to (0,0)$ and $(1,0) \to (0,1)$ are both *possible* at $t_2$
- if at $t_1 + d_{r,\max} - 0$ the automaton is in $(1,0)$, then the transition $(1,0) \to (0,1)$ is *necessary* at $t_2 = t_1 + d_{r,\max}$

b) $i(t_3 - 0) \cdot i(t_3) = 1$ (the input is constant, $i_{|[t_1, t_3]} = 1$)

- if $t_3 \in [t_1, t_1 + d_{r,\min})$ the automaton is in $(1,0)$ necessarily at $t_3$
- if $t_3 \in [t_1 + d_{r,\min}, t_1 + d_{r,\max})$ the transition $(1,0) \to (1,1)$ is *possible* at $t_3$
- if at $t_1 + d_{r,\max} - 0$ the automaton is in $(1,0)$, then the transition $(1,0) \to (1,1)$ is *necessary* at $t_3 = t_1 + d_{r,\max}$
- if $t_3 > t_1 + d_{r,\max}$ the automaton is in $(1,1)$ necessarily at $t_3$.

The rest of the situations are obtained by duality.

**3.2 Remark** The fact that at NIDB the output follows the input with a delay situated somewhere in the intervals $[d_{r,\min}, d_{r,\max}], [d_{f,\min}, d_{f,\max}]$ shows that this automaton is really related with the computation of the identity function on $\boldsymbol{B}$. This fact suggests the next

**3.3 Definition** The states $(0,0)$, $(1,1)$ are called *stable* and the states $(1,0)$, $(0,1)$ are called *unstable*.

**3.4 Remark** In a stable state, under a constant input, the automaton remains indefinitely long:
$$\forall t, \forall t' > t, (i(t) = o(t) \text{ and } i_{|[t,t')} = i(t)) \Rightarrow o_{|[t,t']} = o(t)$$

The automaton does not have this property in an unstable state.

**3.5 Remark** The existence in the behavior of NIDB of some situations of the type
$$Di(t_1) = Di(t_2) = 1 \text{ and } o_{|[t_1, t_2]} = o(t_1)$$
- while the input has switched twice in opposite senses, the output has remained constant- is a property of *inertiality*. We say that the output *filters* the fast switches of the input.

**3.6 Remark** The existence in the behavior of NIDB of the word *possible* is a property of *non-determinism*. Related to this, we have the next special case

**3.7 Definition** If $d_{r,\min} = d_{r,\max} = d_r$ and $d_{f,\min} = d_{f,\max} = d_f$, NIDB is called *deterministic inertial delay buffer* (DIDB).

**3.8 Remark** A DIDB is characterized only by necessary transitions in the following manner:

$$\forall t \geq 0, o(t) = \begin{cases} 1, & \text{if } o(t-0) = 0 \text{ and } i_{|[t-d_r,t)} = 1 \\ 0, & \text{if } o(t-0) = 1 \text{ and } i_{|[t-d_f,t)} = 0 \\ o(t-0), & \text{else} \end{cases}$$

It is clear that
$$\forall t < d_r, o(t) = 0$$

### 4. The Inertial Delay Buffer, Formal Aspects

**4.1 Theorem** Let the signals $i, o$ and the numbers $0 < d_{r,\min} \leq d_{r,\max}, 0 < d_{f,\min} \leq d_{f,\max}$. The next statements, written for $t \geq 0$, are equivalent:

a) $\overline{o(t-0)} \cdot \bigcap_{\xi \in [t-d_{r,\max}, t)} i(\xi) \leq \overline{o(t-0)} \cdot o(t) \leq \overline{o(t-0)} \cdot \bigcap_{\xi \in [t-d_{r,\min}, t)} i(\xi)$

$o(t-0) \cdot \bigcap_{\xi \in [t-d_{f,\max}, t)} \overline{i(\xi)} \leq o(t-0) \cdot \overline{o(t)} \leq o(t-0) \cdot \bigcap_{\xi \in [t-d_{f,\min}, t)} \overline{i(\xi)}$

b) $\overline{o(t-0)} \cdot \bigcap_{\xi \in [t-d_{r,\max}, t)} i(\xi) \cup o(t-0) \cdot \bigcap_{\xi \in [t-d_{f,\max}, t)} \overline{i(\xi)} \leq Do(t) \leq$

$\leq \overline{o(t-0)} \cdot \bigcap_{\xi \in [t-d_{r,\min}, t)} i(\xi) \cup o(t-0) \cdot \bigcap_{\xi \in [t-d_{f,\min}, t)} \overline{i(\xi)}$

**Proof** In the Appendix.

**4.2 Definition** We call NIDB the couple $(i, o)$ of signals so that $\forall t < d_{r,\min}, o(t) = 0$ and one of the conditions 4.1 a), b) is satisfied.

**4.3 Theorem** Let the signals $i, o$ and the numbers $d_r, d_f > 0$. The next statements are equivalent for $t \geq 0$:

a) $\overline{o(t-0)} \cdot o(t) = \overline{o(t-0)} \cdot \bigcap_{\xi \in [t-d_r, t)} i(\xi)$

$o(t-0) \cdot \overline{o(t)} = o(t-0) \cdot \bigcap_{\xi \in [t-d_f, t)} \overline{i(\xi)}$

b) $Do(t) = \overline{o(t-0)} \cdot \bigcap_{\xi \in [t-d_r, t)} i(\xi) \cup o(t-0) \cdot \bigcap_{\xi \in [t-d_f, t)} \overline{i(\xi)}$

c) $\overline{o(t-0)} \cdot \bigcap_{\xi \in [t-d_r, t)} i(\xi) \leq o(t)$

$o(t-0) \cdot \bigcap_{\xi \in [t-d_f, t)} \overline{i(\xi)} \leq \overline{o(t)}$

$\overline{\overline{o(t-0)} \cdot \bigcap_{\xi \in [t-d_r, t)} i(\xi)} \cdot \overline{o(t-0) \cdot \bigcap_{\xi \in [t-d_f, t)} \overline{i(\xi)}} \leq \overline{o(t-0)} \cdot \overline{o(t)} \cup o(t-0) \cdot o(t)$

d) $\overline{o(t-0)} \cdot o(t) \cdot \bigcap_{\xi \in [t-d_r, t)} i(\xi) \cup o(t-0) \cdot \overline{o(t)} \cdot \bigcap_{\xi \in [t-d_f, t)} \overline{i(\xi)} \cup$

$\cup \overline{o(t-0)} \cdot \overline{o(t)} \cdot \overline{\bigcap_{\xi \in [t-d_r, t)} i(\xi)} \cup o(t-0) \cdot o(t) \cdot \overline{\bigcap_{\xi \in [t-d_f, t)} \overline{i(\xi)}} = 1$

**Proof** In the Appendix.

**4.4 Remark** In Theorem 4.3, a), b) repeat 4.1 a), b) in the special case when $d_{r,\min} = d_{r,\max} = d_r$ and $d_{f,\min} = d_{f,\max} = d_f$ and c) repeats 3.8. We read 4.3 d): '*it is true (exactly) one of the next statements made at the time instant $t$:*

- *o switches from 0 to 1 if it was 0 and if $i_{|[t-d_r,t)} = 1$ is true*
- *o switches from 1 to 0 if it was 1 and if $i_{|[t-d_f,t)} = 0$ is true*
- *o is 0 if it was 0 and if $i_{|[t-d_r,t)} = 1$ is not true*
- *o is 1 if it was 1 and if $i_{|[t-d_f,t)} = 0$ is not true*'.

**4.5 Definition** We call DIDB the couple $(i,o)$ of signals so that $\forall t < d_r, o(t) = 0$ and one of the conditions 4.3 a), b), c), d) is satisfied.

**4.6 Lemma** For $t \geq 0$, $d > 0$ the following equations are true

$$\bigcap_{\xi \in [t-d,t)} i(\xi) = i(t-0) \cdot \overline{\bigcup_{\xi \in (t-d,t)} Di(\xi)}$$

$$\bigcap_{\xi \in [t-d,t)} \overline{i(\xi)} = \overline{i(t-0)} \cdot \overline{\bigcup_{\xi \in (t-d,t)} Di(\xi)}$$

**Proof** In the right member of the previous equations we have the property of $i$ of being constant (=continuous) on $(t-d,t)$ in the sense that

$$\forall \xi \in (t-d,t), Di(\xi) = 0$$

thus, by right continuity of $i$ in $t-d$, as signal, we infer that it is constant on $[t-d,t)$ and a value that may be chosen in an arbitrarily point of this interval is $i(t-0)$, respectively $\overline{i(t-0)}$.

**4.7 Remark** We have written equations of the form 4.3 b), see also 4.6

$$Do(t) = \overline{o(t-0)} \cdot i(t-0) \cdot \overline{\bigcup_{\xi \in (t-d,t)} Di(\xi)} \cup o(t-0) \cdot \overline{i(t-0)} \cdot \overline{\bigcup_{\xi \in (t-d,t)} Di(\xi)} =$$

$$= (o(t-0) \oplus i(t-0)) \cdot \overline{\bigcup_{\xi \in (t-d,t)} Di(\xi)}$$

for example in [3], under the generic name of *the equations of the asynchronous automata*. In that context, the Boolean functions to be computed were arbitrary (not the identity, like here). On the other hand, the strong condition of determinism $d_r = d_f = d$ was relaxed by accepting a range of values for $d \in (0,M]$, that becomes parameter, $M$ being a given constant and by the demand that the automaton is stable.

## 5. Comparison with Literature

**5.1 Remark** We rewrite the defining conditions of NIDB from [1] Definition 2, [2] Definition 4 in the spirit and with the notations of this paper under the form

a) $\forall t < d_{r,\min}, o(t) = 0$

b) $\forall t \geq 0, \overline{o(t-0)} \cdot o(t) \leq \bigcup_{t' \in [t-d_{r,\max}, t-d_{r,\min}]} \overline{i(t'-0)} \cdot \bigcap_{\xi \in [t',t)} i(\xi)$

$\forall t \geq 0, o(t-0) \cdot \overline{o(t)} \leq \bigcup_{t' \in [t-d_{f,\max}, t-d_{f,\min}]} i(t'-0) \cdot \bigcap_{\xi \in [t',t)} \overline{i(\xi)}$

c) $\forall t \geq 0, \overline{i(t-0)} \cdot i(t) \leq \bigcup_{t' \in (t, t+d_{r,\max})} i(t'-0) \cdot \overline{i(t')} \cup \bigcup_{t' \in [t+d_{r,\min}, t+d_{r,\max}]} \overline{o(t'-0)} \cdot o(t')$

$$\forall t \geq 0, \overline{i(t-0)} \cdot \overline{i(t)} \leq \bigcup_{t' \in (t, t+d_{f,\max})} \overline{i(t'-0)} \cdot i(t') \cup \bigcup_{t' \in [t+d_{f,\min}, t+d_{f,\max}]} \overline{o(t'-0)} \cdot o(t')$$

We mention that in the case of b) (like in other similar situations from this paper) the fact that $t$ runs in $[0, \infty)$ does not contradict the sense of these implications because, for example, the 'event' $\overline{o(t-0)} \cdot o(t) = 1$ is not possible if $t < d_{r,\min}$ etc.

**5.2 Theorem** 4.1 a) implies 5.1 b).
**Proof** In the Appendix.

**5.3 Counterexample** showing that 5.1 b) does not imply 4.1 a):
$$i(t) = \begin{cases} 1, \text{if } t \geq 0 \\ 0, \text{else} \end{cases}$$
$$\forall t, o(t) = 0$$

5.1 b) is true, but 4.1 a) is false at $t = d_{r,\max}$.

**5.4 Counterexample** showing that 4.1 a) does not imply 5.1 c). For $d_{r,\min} = d_{r,\max} = d_{f,\min} = d_{f,\max} = 2$ and
$$i(t) = \begin{cases} 1, \text{if } t \in [1,2) \\ 0, \text{else} \end{cases}$$
$$\forall t, o(t) = 0$$

4.1 a) is satisfied, but for $t = 2$, 5.1 c) is false:
$$\overline{i(2-0)} \cdot i(2) = 1 \text{ and } \forall t' \in (2,4), \overline{i(t'-0)} \cdot i(t') = 0 \text{ and } \overline{o(4-0)} \cdot o(4) = 0$$

**5.5 Remark** Because the couple $i, o$ from 5.4 agrees with our intuition, as stated in section 3, the conclusion is that 5.1 is incorrect. Our opinion is that in [1], [2] the definition of NIDB is incorrect.

<div align="center">

**Appendix**

</div>

**A.1 The proof of Theorem 4.1.**
The two left implications of a):
$$\overline{o(t-0)} \cdot \bigcap_{\xi \in [t-d_{r,\max}, t)} \overline{i(\xi)} \cup \overline{o(t-0)} \cdot o(t) = 1$$

$$\overline{o(t-0) \cdot \overline{\bigcap_{\xi \in [t-d_{f,\max},t)} \overline{i(\xi)}} \cup o(t-0) \cdot \overline{o(t)}} = 1$$

are equivalent with their product, i.e.

$$1 = (o(t-0) \cup \overline{\bigcap_{\xi \in [t-d_{r,\max},t)} i(\xi)} \cup \overline{o(t-0) \cdot o(t)}) \cdot (\overline{o(t-0)} \cup \bigcap_{\xi \in [t-d_{f,\max},t)} \overline{i(\xi)} \cup o(t-0) \cdot \overline{o(t)}) =$$

$$= o(t-0) \cdot \bigcap_{\xi \in [t-d_{f,\max},t)} \overline{i(\xi)} \cup o(t-0) \cdot \overline{o(t)} \cup \overline{o(t-0)} \cdot \overline{\bigcap_{\xi \in [t-d_{r,\max},t)} i(\xi)} \cup$$

$$\cup \overline{\bigcap_{\xi \in [t-d_{r,\max},t)} i(\xi)} \cdot \bigcap_{\xi \in [t-d_{f,\max},t)} \overline{i(\xi)} \cup o(t-0) \cdot \overline{o(t)} \cdot \overline{\bigcap_{\xi \in [t-d_{r,\max},t)} i(\xi)} \cup$$

$$\cup \overline{o(t-0) \cdot o(t)} \cup \overline{o(t-0) \cdot o(t)} \cdot \bigcap_{\xi \in [t-d_{f,\max},t)} \overline{i(\xi)} =$$

$$= (o(t-0) \cdot \bigcap_{\xi \in [t-d_{f,\max},t)} \overline{i(\xi)} \cup \overline{o(t-0)} \cdot \overline{\bigcap_{\xi \in [t-d_{r,\max},t)} i(\xi)} \cup \overline{\bigcap_{\xi \in [t-d_{r,\max},t)} i(\xi)} \cdot \bigcap_{\xi \in [t-d_{f,\max},t)} \overline{i(\xi)}) \cup$$

$$\cup (\overline{o(t-0) \cdot o(t)} \cup o(t-0) \cdot \overline{o(t)}) =$$

$$= \overline{(\overline{o(t-0)} \cup \bigcap_{\xi \in [t-d_{f,\max},t)} \overline{i(\xi)})} \cdot (o(t-0) \cup \overline{\bigcap_{\xi \in [t-d_{r,\max},t)} i(\xi)}) \cdot (\overline{\bigcap_{\xi \in [t-d_{r,\max},t)} i(\xi)} \cup \bigcap_{\xi \in [t-d_{f,\max},t)} \overline{i(\xi)}) \cup$$

$$\cup Do(t) =$$

$$= \overline{(\overline{o(t-0)} \cdot \bigcap_{\xi \in [t-d_{r,\max},t)} i(\xi) \cup o(t-0) \cdot \bigcap_{\xi \in [t-d_{f,\max},t)} \overline{i(\xi)}) \cdot (\overline{\bigcap_{\xi \in [t-d_{r,\max},t)} i(\xi)} \cup \bigcap_{\xi \in [t-d_{f,\max},t)} \overline{i(\xi)})} \cup$$

$$\cup Do(t) =$$

$$= \overline{(\overline{o(t-0)} \cdot \bigcap_{\xi \in [t-d_{r,\max},t)} i(\xi) \cup o(t-0) \cdot \bigcap_{\xi \in [t-d_{f,\max},t)} \overline{i(\xi)})} \cup Do(t)$$

The proof is similar for the other two right implications of a).

## A.2 The proof of Theorem 4.3.

a) $\Leftrightarrow$ b) was proved at 4.1 and b) $\Leftrightarrow$ d), c) $\Leftrightarrow$ d) are proved by direct computation. We show b) $\Leftrightarrow$ d).

$$1 = (\overline{o(t-0) \cdot o(t)} \cup \overline{o(t-0) \cdot \overline{o(t)}} \cup \overline{o(t-0)} \cdot \bigcap_{\xi \in [t-d_r,t)} i(\xi) \cup o(t-0) \cdot \bigcap_{\xi \in [t-d_f,t)} \overline{i(\xi)}) \cdot$$

$$\cdot (\overline{o(t-0) \cdot o(t)} \cup o(t-0) \cdot \overline{o(t)} \cup (o(t-0) \cup \overline{\bigcap_{\xi \in [t-d_r,t)} i(\xi)}) \cdot (\overline{o(t-0)} \cup \bigcap_{\xi \in [t-d_f,t)} \overline{i(\xi)})) =$$

$$= (\overline{o(t-0) \cdot o(t)} \cup o(t-0) \cdot \overline{o(t)} \cup \overline{o(t-0)} \cdot \bigcap_{\xi \in [t-d_r,t)} i(\xi) \cup o(t-0) \cdot \bigcap_{\xi \in [t-d_f,t)} \overline{i(\xi)}) \cdot$$

$$\cdot (\overline{o(t-0) \cdot o(t)} \cup o(t-0) \cdot \overline{o(t)} \cup o(t-0) \cdot \bigcap_{\xi \in [t-d_f,t)} \overline{i(\xi)} \cup \overline{o(t-0)} \cdot \overline{\bigcap_{\xi \in [t-d_r,t)} i(\xi)} \cup$$

$$\cup \overline{\bigcap_{\xi \in [t-d_r,t)} i(\xi)} \cdot \bigcap_{\xi \in [t-d_f,t)} \overline{i(\xi)}) =$$

$$= \overline{o(t-0) \cdot o(t)} \cdot \bigcap_{\xi \in [t-d_r,t)} i(\xi) \cup \overline{o(t-0) \cdot o(t)} \cdot \overline{\bigcap_{\xi \in [t-d_r,t)} i(\xi)} \cdot \bigcap_{\xi \in [t-d_f,t)} \overline{i(\xi)} \cup$$

$$\cup \overline{o(t-0) \cdot o(t) \cdot \overline{\bigcap_{\xi \in [t-d_f,t)} i(\xi)}} \cup \overline{o(t-0) \cdot o(t) \cdot \overline{\bigcap_{\xi \in [t-d_r,t)} i(\xi)} \cdot \overline{\bigcap_{\xi \in [t-d_f,t)} i(\xi)}} \cup$$

$$\cup \overline{\overline{o(t-0)} \cdot o(t) \cdot \bigcap_{\xi \in [t-d_r,t)} i(\xi)} \cup \overline{o(t-0) \cdot \overline{o(t)} \cdot \overline{\bigcap_{\xi \in [t-d_f,t)} i(\xi)}}$$

$$= \overline{o(t-0) \cdot o(t) \cdot \overline{\bigcap_{\xi \in [t-d_r,t)} i(\xi)}} \cup \overline{o(t-0) \cdot o(t) \cdot \overline{\bigcap_{\xi \in [t-d_f,t)} i(\xi)}} \cup$$

$$\cup \overline{\overline{o(t-0)} \cdot o(t) \cdot \bigcap_{\xi \in [t-d_r,t)} i(\xi)} \cup \overline{o(t-0) \cdot \overline{o(t)} \cdot \overline{\bigcap_{\xi \in [t-d_f,t)} i(\xi)}}$$

### A.3 The proof of Theorem 5.2

We show that if 4.1 a) is true and 5.1 b) is false, we get a contradiction. The hypothesis states:

$$\exists t \geq 0, \overline{o(t-0)} \cdot o(t) = 1 \text{ and } \forall t' \in [t-d_{r,\max}, t-d_{r,\min}], \overline{i(t'-0)} \cdot \bigcap_{\xi \in [t',t)} i(\xi) = 0$$

By taking into account 4.1 a) also

$$\bigcap_{\xi \in [t-d_{r,\min},t)} i(\xi) = 1 \text{ and } i(t-d_{r,\min}-0) = 1$$

If $(t_k)$ is a timed sequence making 2.5 b) true for $i$, a finite number of $k's$ exists for which $t_k \in [t-d_{r,\max}, t-d_{r,\min}]$ and

$$i(t_k) = i(t_k - 0) = 1$$

and eventually

$$i(t-d_{r,\max}) = i(t-d_{r,\max} - 0) = 1$$

But the semi-derivative $\overline{o(t-0)} \cdot o(t)$ must be null at the left of $t$ (the derivative of a signal may be 1 only in the discrete points of a timed sequence) and the first inequality 4.1 a) from the left gives the contradiction

$$\overline{o((t-0)-0)} \cdot \bigcap_{\xi \in [t-d_{r,\max}-0, t-0)} i(\xi) = \overline{o(t-0)} \cdot i(t-d_{r,\max}-0) \cdot \bigcap_{\xi \in [t-d_{r,\max},t)} i(\xi) = 1 \leq$$

$$\leq 0 = \overline{o((t-0)-0)} \cdot o(t-0) = \overline{o(t-0)} \cdot o(t-0)$$

(The last statements contain some unproved 'almost obvious' facts that belong rather to mathematical analysis than to this context, but they are easily accepted by the reader, we hope).